\documentclass[sigconf]{acmart}

\usepackage{graphicx}
\usepackage[utf8]{inputenc}
\usepackage{amsmath}
\AtBeginDocument{%
  \providecommand\BibTeX{{%
    \normalfont B\kern-0.5em{\scshape i\kern-0.25em b}\kern-0.8em\TeX}}}

\acmYear{2020}
\copyrightyear{2020}
\setcopyright{rightsretained}
\acmConference[FAT* '20]{Conference on Fairness, Accountability, and Transparency}{January 27--30, 2020}{Barcelona, Spain}
\acmBooktitle{Conference on Fairness, Accountability, and Transparency (FAT* '20), January 27--30, 2020, Barcelona, Spain}
\acmPrice{}
\acmDOI{10.1145/3351095.3372849}
\acmISBN{978-1-4503-6936-7/20/01}

\begin{document}

\title[What does it mean to `solve' the problem of discrimination in hiring?]{What does it mean to `solve' the problem of discrimination in hiring? Social, technical and legal perspectives from the UK on automated hiring systems}
\author{Javier S\'anchez-Monedero}
\email{Sanchez-monederoJ@cardiff.ac.uk}
\orcid{0001-8649-1709}
\affiliation{%
  \institution{Cardiff University}
  \streetaddress{Two Central Square, Central Square}
  \city{Cardiff}
  \state{Wales}
  \country{United Kingdom}
  \postcode{CF10 1FS}
}

\author{Lina Dencik}
\email{DencikL@cardiff.ac.uk}
\affiliation{%
  \institution{Cardiff University}
  \streetaddress{Two Central Square, Central Square}
  \city{Cardiff}
  \state{Wales}
  \country{United Kingdom}
  \postcode{CF10 1FS}
}

\author{Lilian Edwards}
\email{lilian.edwards@newcastle.ac.uk}
\affiliation{%
  \institution{University of Newcastle}
  \streetaddress{21-24 Windsor Terrace}
  \city{Newcastle upon Tyne}
  \state{England}
  \country{United Kingdom}
  \postcode{NE1 7RU}
}

\renewcommand{\shortauthors}{Javier S\'anchez-Monedero, Lina Dencik, and Lilian Edwards}

\begin{abstract}
  Discriminatory practices in recruitment and hiring are an ongoing issue
  that is a concern not just for workplace relations, but also for wider
  understandings of economic justice and inequality. The ability to get
  and keep a job is a key aspect of participating in society and
  sustaining livelihoods. Yet the way decisions are made on who is
  eligible for jobs, and why, are rapidly changing with the advent and
  growth in uptake of automated hiring systems (AHSs) powered by
  data-driven tools. Evidence of the extent of this uptake around the
  globe is scarce, but a recent report estimated that 98\% of Fortune 500
  companies use Applicant Tracking Systems of some kind in their hiring
  process, a trend driven by perceived efficiency measures and
  cost-savings. Key concerns about such AHSs include the lack of
  transparency and potential limitation of access to jobs for specific
  profiles. In relation to the latter, however, several of these AHSs
  claim to detect and mitigate discriminatory practices against protected
  groups and promote diversity and inclusion at work. Yet whilst these
  tools have a growing user-base around the world, such claims of `bias
  mitigation' are rarely scrutinised and evaluated, and when done so, have
  almost exclusively been from a US socio-legal perspective.
  
  In this paper, we introduce a perspective outside the US by critically
  examining how three prominent automated hiring systems (AHSs) in regular
  use in the UK, HireVue, Pymetrics and Applied, understand and attempt to
  mitigate bias and discrimination. These systems have been chosen as they
  explicitly claim to address issues of discrimination in hiring and,
  unlike many of their competitors, provide some information about how
  their systems work that can inform an analysis. Using publicly available
  documents, we describe how their tools are designed, validated and
  audited for bias, highlighting assumptions and limitations, before
  situating these in the socio-legal context of the UK. The UK has a very
  different legal background to the US in terms not only of hiring and
  equality law, but also in terms of data protection (DP) law. We argue
  that this might be important for addressing concerns about transparency
  and could mean a challenge to building bias mitigation into AHSs
  definitively capable of meeting EU legal standards. This is significant
  as these AHSs, especially those developed in the US, may obscure rather
  than improve systemic discrimination in the workplace.
\end{abstract}

\ccsdesc[500]{Social and professional topics~Socio-technical systems}
\ccsdesc[300]{Social and professional topics~Systems analysis and design}
\ccsdesc[300]{Applied computing~Law}
\ccsdesc[300]{Applied computing~Sociology}

\keywords{Socio-technical systems, automated hiring, algorithmic decision-making, fairness, discrimination, GDPR, social justice}


\maketitle

\section{Introduction}

The use of data systems and automated decision-making as a way to
monitor, allocate, assess and manage labour is a growing feature of the
contemporary workplace. Of increasing significance is the way Human
Resources, and hiring practices in particular, are being transformed
through various forms of automation and the use of data-driven
technologies which we will collectively term Automated Hiring Systems
(AHSs) \cite{bogen_help_2018,sam_datafication_2018,sanchez-monedero_datafication_2019}. Whilst there is a lack of data on the global
uptake of such technologies, a recent report estimated that 98\% of
Fortune 500 companies use Applicant Tracking Systems of some kind in
their hiring process \cite{shields_over_2018}. The so-called `hiring funnel' \cite{bogen_help_2018}
consists of sourcing, screening, interviewing, and selection/rejection
as a set of progressive filtering stages to identify and recruit the
most suitable candidates. Each of these stages are undergoing forms of
automation, as part of not only perceived efficiency measures and
cost-savings that data-driven technologies afford, but also as a means
to detect and mitigate discriminatory practices against protected groups
and promoting diversity and inclusion at work \cite{bogen_help_2018}. Hiring platforms
such as PeopleStrong or TribePad implement basic measures to mitigate
human unconscious biases, such as anonymisation of candidates, while
other platforms such as HireVue, Pymetrics and Applied\footnote{Note
  that Applied does not automate the evaluation of candidates but it
  assists in the process of bias discovery and mitigation.} claim to
specifically tackle the problem of discrimination in hiring. Yet the
basis upon which such claims are made is rarely scrutinised and
evaluated. While algorithmic bias generally and in employment law
specifically \cite{barocas_solon_big_2016} has had extensive investigation in the FAT*
community, literature on bias mitigation in AHSs is at an early stage
and so far primarily focused on the US context of employment both
socially and legally \cite{ajunwa_ifeoma_platforms_2019,bogen_help_2018,raghavan_mitigating_2020}. We know much less about how this
phenomena is developing in Europe \cite{algorithmwatch_automating_2019}. This is the first scholarly
attempt to consider the question of how satisfactorily bias and
discrimination might be mitigated in AHSs within the UK
context.\footnote{In this paper we are not looking at the general
  scientific validation of the system design as we do not have access to
  any independent studies of that kind. Instead, we are focusing on how
  `bias' and discrimination is understood in the design. Our assessment
  does therefore not directly engage with the effectiveness of using
  AHSs to evaluate candidates, even though issues of discrimination
  would be very pertinent in such an assessment.}

We start by outlining how data-driven technologies are transforming
hiring practices, before turning to focus on three AHSs widely in use in
the UK which make claims to deal with bias and whose claims could be
evaluated using publicly available materials such as company white
papers and reports, patents, marketing resources, seminars and, in one
case, source code. Access to further information relating to code, data
sets, features design, trained models, or even the application user
interface was not possible, and will often vary depending on client.
Based on this publicly available material that outline their
technological and procedural frameworks, we examine how these products
implement bias discovery and mitigation. In doing so, and in building on
recent work in this area \cite{ajunwa_paradox_2020,bogen_help_2018,raghavan_mitigating_2020,sanchez-monedero_datafication_2019} we explore the assumptions
made about the meaning of bias and discrimination in hiring practices
embedded within these tools. The three prominent systems we examined
were Pymetrics, HireVue and Applied. For each of these, we outline the
design of the tool, focusing on how bias in hiring is addressed. The
first two of these come from the US, and the third, Applied, from the
UK, for comparative purposes. Other UK systems were considered but
either made no claims as to debiasing, or did not make available
sufficient public material to analyse. Although not rigorously
addressed, this gap, either in implementation of debiasing, or its
disclosure, is in itself, we feel, a significant finding deserving
further research.

Next we briefly discuss the background to UK and EU equality law in the
context of employment. Importantly, we raise the point that in the EU,
all natural persons whose personal data is processed are given data
rights, including rights to transparency and protection against power
imbalance, which may be as useful in combating bias in AHSs as
employment equality rights; these data protection (DP) rights, ensconced
in the EU General Data Protection Regulation (GDPR), are not found in
omnibus fashion in the US and thus have largely not been analysed in the
US hiring algorithm literature.

In analysing these systems, we started from the premise that there is no
singular or unified way of interpreting the meaning of discrimination,
or how it might feature in hiring practices, nor is there consensus on
any computational criteria for how ``bias'' should be defined, made
explicit, or mitigated \cite{friedler_comparative_2019,overdorf_questioning_2018,raghavan_mitigating_2020,sanchez-monedero_how_2018,verma_fairness_2018}. Ongoing analysis point out
issues such as the unsuitability of maths to capture the full meaning of
social concepts such as fairness, especially in a general sense, or the
risk of technological solutionism \cite{selbst_fairness_2019}. Even when considering one
statistical definition for bias such as the error rate balance amongst
groups, the understanding and implication of that approach radically
varies with the context and the consequential decisions that are driven
by the algorithmic output \cite{hellman_measuring_2019}. Also, all sociotechnical systems,
even when designed to mitigate biases, are designed with use cases in
mind that may not hold in all scenarios \cite{selbst_fairness_2019}. Moreover, fairness can
be procedural, as in the equal treatment of individuals, or substantive,
as in the equal impact on groups \cite{barocas_solon_big_2016}, what is also referred to as
opportunity-based vs. outcome-based notions of bias. These do not
necessarily align and may actually be contradictory \cite{lipton_does_2018}.

We conclude by arguing that, while common practices may be emerging to
\emph{mitigate} bias in AHSs built in the US, these are inevitably
likely to reflect US legal and societal conceptions of bias and
discrimination, and yet are exported wholesale as products to UK and
other markets beyond the US. If this mismatch is not explicitly
disclosed and analysed, there is a patent danger that inappropriate
US-centric values and laws relating to bias in hiring (and more
generally in society) may be exported to UK workplaces. Data protection
rights, such as the alleged ``right to an explanation'' may by contrast
not be implemented, potentially rendering useless the rights of
candidates. What is more, while biased workplace hiring practices in
traditional modes of hiring may at least to some extent be evident, and
combatted in traditional ways by unions, strategic litigation and
regulators, there are severe dangers that such biases buried within
``black box'' AHSs may not be manifestly obvious and so remain
unchallenged. As AHSs become ever more popular, especially in sectors
which are precarious, such as retail and the gig economy where
prospective employees (or contractors) may have little economic power or
knowledge of rights \cite{ajunwa_ifeoma_platforms_2019,bogen_help_2018}, this is, we argue, a serious problem.

\section{The data-driven hiring funnel}

The automation of hiring is an important part of the broader discussion
on the future of work, subject to a range of ethical concerns and issues
of fairness across the different stages of the hiring funnel, from
sourcing to screening to interviewing to selection \cite{bogen_help_2018}. At each of
these stages, the increasing use of data to automate or partially
automate the process is significantly changing the way decisions are
made on who is eligible for jobs, and why. Recommender engines based on
hybrid collaborative filtering methods are used to capture user (both
job seekers and recruiters) preferences; tools are used to filter
candidates and identify the most promising candidates; automated skills
tests and video interviews evaluate candidates; and analytics dashboards
are used to select candidates and generate ad hoc job offers \cite{sanchez-monedero_datafication_2019}.
Data is collected from a range of sources and can be self-reported by
the candidate in the form of unstructured documents such as resumés or
as structured professional network profiles or online application forms.
Often this information will be extended and/or scored through additional
sources of information and assessment tests.

Central to the hiring process, only made more salient with automation,
is the goal of hiring for `fit' with an organisation, a criterion that
leverages the employer significant discretion and may only be useful if
an employer can make an accurate determination of such a fit \cite{ajunwa_paradox_2020}.
The algorithmic specification of such a fit often relies on abstracting
candidate profiles in relation to historical data of the company and/or
top performers for specific roles. That is, at each stage of the hiring
process, tools are designed around a set of variables, optimized for a
particular criteria as to who might constitute an appropriate and `good'
employee. Deciding on such a criteria has long been imbued with the
potential for discriminatory practices, and the aim here is not to
suggest that those practices have only emerged with the introduction of
data-driven tools nor is it to further the construct of a (false) binary
between human vs. automated decision-making. Rather, in line with Ajunwa
\cite{ajunwa_paradox_2020}, we see value in addressing issues that such tools have made
salient with the understanding that technologies are the product of
human action and that, as Wajcman puts it, ``histories of discrimination
live on in digital platforms and become part of the logic of everyday
algorithmic systems.'' \cite{wajcman_automation:_2017}.

At the same time, the turn to AHSs is significant also for the potential
scale of impact, the difficulty with interrogating decision-making,
increasing information asymmetry between labour and management, the
standardisation of techniques, the obfuscation of accountability, and
the veneer of objectivity that such technologies often afford employers,
despite evidence of discrimination at each stage of the data-driven
hiring funnel. For instance, when sourcing, specific groups have been
found excluded from viewing job ads based on age \cite{angwin_dozens_2017} or gender
\cite{ariana_tobin_facebook_2018}. Masculine gendered language in job descriptions can discourage
women to apply for certain type of jobs \cite{gaucher_evidence_2011} whilst language used by
candidates has been demonstrated to be a proxy for social class that
impact on the chances of being selected for an interview \cite{thomas_labor_2018}. As
such, screening of candidates based on collaborative filtering can
perpetuate existing discriminatory practices \cite{bogen_help_2018}. Even when a
recommendation engine is generally not discriminating against women,
some job titles and rank results can place men in better positions
\cite{chen_investigating_2018}. Automated interviews, including the use of speech and facial
recognition software, has been shown to perform poorly, particularly
with regards to women of colour \cite{bogen_help_2018,buolamwini_gender_2018}. Moreover, using `cultural
fit' as part of the selection criterion has been shown to lead to
exclusive hiring practices, and is now outlawed by some companies as
part of an effort to decrease unconscious bias \cite{ajunwa_paradox_2020}. These findings
are part of a broader debate that engages with the way data-driven
technologies systematically introduce and entrench forms of
discrimination and social and economic inequality \cite{angwin_machine_2016,noble_algorithms_2018,oneil_weapons_2016,sanchez-monedero_how_2018,pena_gangadharan_data_2014}.

The role that AHSs come to play in entrenching and furthering the
information asymmetry between labour and management inherent in the
hiring process, the increased lack of accountability in decision-making,
and the advancement of a global standard of management techniques
changes the terms of control in the workplace; what Ajunwa \cite{ajunwa_rise_2018}
refers to as `platform authoritarianism' in which employers gain
penetrating new insights into current or potential employees, but the
latter have no room to negotiate. What is more, the reliance on what are
often candidate profiles based on incomplete data, proxies and
inferences, may further this sense of authoritarianism on the basis of
incomplete or inaccurate profiles. Similarly, Moore \cite{moore_quantified_2016} argues that
the on-going quantification of the workplace comes to discipline workers
as they continuously seek to adapt to the needs of the technologies in
place to assess them in a process of `self-quantification'. Importantly,
who might be best positioned to adapt to such measures and who is likely
to be excluded rarely forms part of approaches to bias mitigation.
Instead, the focus of mitigation tends to be on the technicalities of
the model, at the point of the interface, situating the relationship
between discrimination and inequality within the confines of
`unconscious bias'.

Of course, asking providers of AHSs to attend to the dynamics of power
in labour relations and society more broadly might seem unnecessarily
burdensome, but by not recognising the broader functions of automation
in shaping those dynamics when considering forms of bias mitigation, we
run the risk of neutralising challenges in a way that actively
facilitates discrimination and inequality under a banner of fairness
\cite{dencik_conceptual_2018,gangadharan_decentering_2019,hoffmann_where_2019}. This is particularly important as these companies are
part of standardising not only managerial techniques \cite{ajunwa_ifeoma_platforms_2019}, but also
how we should both understand and `solve' the problem of discrimination
in hiring, within a potentially global market.

\section{Technological bias auditing and mitigation in hiring}

The data-driven hiring funnel therefore demands attention and scrutiny,
particularly in relation to issues of discrimination and rights. This is
especially pertinent as a number of AHSs specifically claim to address
bias and discrimination in hiring, and seek to do so across
organisational and national contexts. The claim is that automation
reduces hirer bias, replacing messy human decisions with a neutral
technical process \cite{ajunwa_ifeoma_platforms_2019}. In this section, we introduce three such
software systems that specifically address the issue of bias in hiring.
These AHSs were selected as they are known to be used in the UK, and,
unlike many of their competitors, there is publicly available
information to inform an evaluation of their approach. We base this
evaluation on what documents are available through their websites and
registered patents. The complete models or design of the tools, or
outlines of specific data sources, have not been available to us for
auditing and will vary depending on the client.

\subsection{Pymetrics}

\begin{figure}[h]
  \centering
  \includegraphics[width=\linewidth]{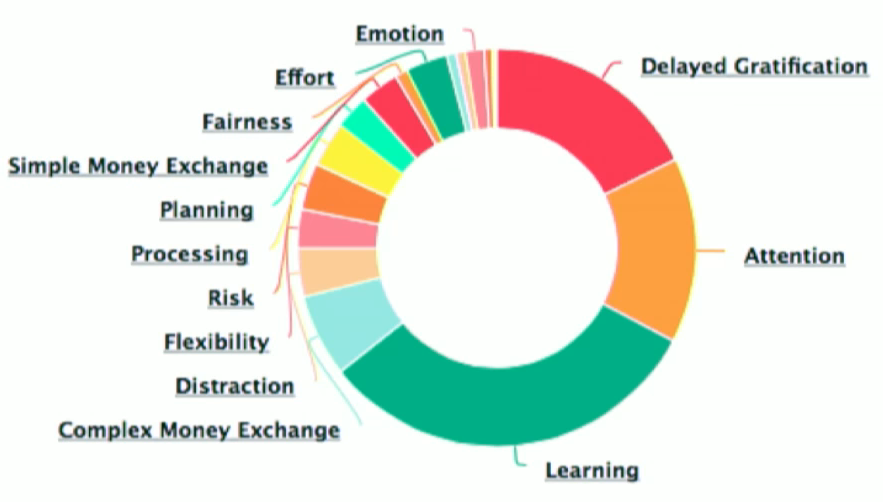}
  \caption{Pymetrics software engineer profile of traits.
  Screenshot from Pymetrics seminar recording available in \cite{yoo_pymetrics_2017}.}
  \label{fig:pymetrics}
\end{figure}

Pymetrics is a vendor of hiring technology that performs a
pre-employment assessment of candidates with games tests that are based
on neuroscience research\footnote{\href{https://www.pymetrics.com/}{{https://www.pymetrics.com/}}}.
By analysing how participants behave with these games, the software
generates metrics of cognitive, social and emotional traits. The profile
is evaluated with a statistical model trained on the game results of top
performers in each role so that the model can calculate a score and
categorize the candidate as out-of-group and in-group \cite{polli_systems_2019}. This
fit-to-role score is an aggregation of the scores of the individual
tests. To create a description of optimal traits values for a
professional role, the system uses an unsupervised learning clustering
algorithm to identify representative scores of traits for the reference
workers. For instance, according to Pymetrics \cite{yoo_pymetrics_2017}, one of the
desired characteristics of the best performing software developers are
`delayed gratification' and `learning', that are two of the
characteristics of a person that the software can measure (see Figure \ref{fig:pymetrics}). A person with a good score in these characteristics is more likely
to fit in the position.

Pymetrics features bias mitigation in candidate assessment based on the
game design that is intended to avoid score correlation with protected
groups and be agnostic with respect to non-verbal communication and
culture \cite{yoo_pymetrics_2017}. 
To evaluate the fairness of each game score and of the
fit-to-role score, Pymetrics performs statistical tests to compare both
the individual and aggregated scores with respect to candidates grouped
by age, gender and ethnicity. Since each game produces one to ten
metrics, each one of those metrics is analysed to check for negative
impact on protected groups scores.

The Pymetrics's US patent \cite{polli_systems_2019} reports detailed statistical test
results of multiple group comparison. The worker and candidate data used
for the test corresponds to several Pymetrics customers. The impact of
age was analysed by grouping candidates in four age groups ($\leq$ 29; 30-34;
35-39; and $\geq$ 40), the impact on gender was evaluated considering binary
gender classes (male; female) and the impact of race considered eight
categories (Asian, Black, Hispanic, Middle Eastern, Native American,
White, other, and mixed race). The multivariate ANOVA and Hotelling's
T-squared tests concluded that for each game none of the tasks showed
significant differences by ethnic group and a subset of the tasks showed
different scores based on age and gender. When comparing the final score
that assesses the suitability of a person for a position, the
multivariate statistical analyses concluded that there were no
differences, statistical bias, between samples groups by age
(p$>$0.05), gender (p$>$0.05) and race
(p$>>$0.1).

Pymetrics developed a specific tool, audit-AI, to perform this auditing
and released it as open source software\footnote{\href{https://github.com/pymetrics/audit-ai}{{https://github.com/pymetrics/audit-ai}}}.
The software also performs US regulatory compliance checks to comply to
fair standard treatment of all protected groups indicated in the Uniform
Guidelines on Employee Selection Procedures by the Equal Employment
Opportunity Commission (EEOC) \cite{us_eeoc_adoption_1979}. The EEOC requires that the ratio
of the proportion of pass rates, selection of candidates in this
context, of the highest-passing and lowest-passing demographic
categories has to follow the 4/5ths rule, meaning the ratio comparing
the two extreme cases cannot be smaller than 0.80. For example, if there
are 1,000 candidates who were hired and they belong to three groups, A,
B and C, with passing frequency of 350, 320 and 330 respectively, the
highest and lowest passing groups are A and B and, so the bias ratio is
320/350, or 0.91. Since this ratio is greater than 0.80 the selection
procedure meets the legal requirements of the EEOC.

\subsection{HireVue}

HireVue\footnote{\href{https://www.hirevue.com/}{{https://www.hirevue.com/}}}
is a product to automate the pre-interview assessment of candidates from
a pool. It performs automated video interview and games to profile
candidates. The games and questions are designed based on Industrial
Organization psychology research. The tool extracts three types of
\emph{indicator} data from applicants: categorical, audio and video
\cite{taylor_model-driven_2017}. HireVue promises to eliminate human biases in the assessment
of candidates whilst simultaneously finding the subset of candidates
that are most likely to be successful in a job by comparing them to
employees already performing that job. Therefore, the software automates
the screening of candidates by directly selecting them for a later human
interview. In contrast to information provided by Pymetrics, HireVue
provides more general information about what precise features are
extracted from candidates and the specific statistical definition of
bias.

Bias detection is performed by measuring demographic parity as defined
by the US EEOC. To mitigate bias, HireVue has two strategies:

The first strategy consists of the removal of indicators that have an
adverse impact on protected groups based on previous knowledge. As
described in the website documentation \cite{hirevue_bias_2019} and patent \cite{taylor_model-driven_2017}, the
abstract process first defines performance indicators and questions to
elicit responses that can be measured and related to job performance.
Indicators can include not only what the candidate says but also how
they say it by extracting audio features such as pitch or duration. Then
a model is trained to learn how to predict the suitability of the
candidate from all these indicators. The bias mitigation consists of
evaluating the adverse impact on protected groups by detecting violation
of the 4/5ths rule. Features that cause biased results are removed and
the models are re-trained and re-evaluated until bias is not detected.
As an illustrative example, HireVue presents the case of speaking slowly
as a characteristic of the top performers in a technical support role
that also is more common in men\footnote{https://www.hirevue.com/blog/hirevue-assessments-and-preventing-algorithmic-bias}.
Testing the tool should reveal this correlation so that the feature will
be suppressed from the model input\footnote{HireVue does not specify how
  this correlation can be discovered.}. Alternatively, according to the
patent \cite{taylor_model-driven_2017}, the bias discovering process can consist of applying
clustering methods to detect protected groups in the feature
space\footnote{Here we refer to as \emph{feature space} to input data
  that consist of features extracted from candidates.}. Clustering
methods are unsupervised learning algorithms that try to structure
unlabelled data points into different groups based on their arrangement
in the input space. In this case, data points are composed of indicators
excluding those ones related to group and performance. The proposal
consists of running these methods to try to find groups based on the
features used to evaluate candidates. If the method is capable of
discovering protected groups in this unsupervised manner, this means
that one or more features, such as weight or hair colour, are correlated
with the categorical variable so that learning algorithms could
potentially use these features to learn to discriminate. If this is the
case, the input data will be examined to identify and remove such
correlated features.

The second strategy consists of modifying the learning algorithm to
account for fairness. In machine learning, the objective function is a
mathematical expression of how well the model is fitted to the data. It
guides the learning algorithms in the process of learning from data and
creating data transformations that contribute to improving accuracy. The
patent \cite{larsen_performance_2017} proposes to replace the objective function, typically a
global sum of squared errors, with a corrected function that sums the
separate error of the model for each protected group. By doing so, the
objective function incorporates a fairness constraint that will
indirectly introduce pressure on the learning algorithm to build a model
that considers that the accuracy of the model with respect to all the
protected groups (race, gender, age, etc.) must be equal. To account for
the equal influence of underrepresented or minority classes, each group
error term is normalized to ensure that the majority class does not
influence the model more than the rest of the classes. The general
expression for corrected error can be written as:
\begin{displaymath}
  E_{\text{corrected}} = E_{A} + E_{B} + E_{C} + \cdots
\end{displaymath}

where $E_{A}$, $E_{B}$, etc. are the errors for
each protected group.

Additionally, the patent proposes to sum a penalty term to the corrected
error to account for the regulations such as the EEOC. An example term
of the 4/5ths rule can be represented as follows:
\begin{displaymath}
  P\left( X \right) = \begin{cases} p_{m} &\mbox{if } f\left( X \right) \mbox{ violates 4/5ths rule} \\
  0 & \mbox{otherwise }\end{cases} 
\end{displaymath}

where $p_{m}$ is the cost the user wants to associate with the
rule violation and $f\left( X \right)$ refers to the candidate
evaluation model whose output will be checked for demographic parity.
Therefore, the objective error function becomes:
\begin{displaymath}
  E_{\text{with\_ penalty}} = E_{\text{corrected}} + P\left( X \right)
\end{displaymath}

\subsection{Applied}

Applied\footnote{\href{https://www.beapplied.com}{{https://www.beapplied.com}}}
is a hiring platform specialised in promoting diversity and inclusion in
recruitment. The system includes a numerical, analytical and
problem-solving testing platform called Mapped\footnote{\href{http://www.get-mapped.com/}{{http://www.get-mapped.com/}}}
that designs the tests by excluding patterns that are found to
negatively impact on different demographic groups and improve pass rates
of candidates of different groups\footnote{\href{http://www.get-mapped.com/\#about}{{http://www.get-mapped.com/\#about}}}.

\begin{figure}[h]
  \centering
  \includegraphics[width=\linewidth]{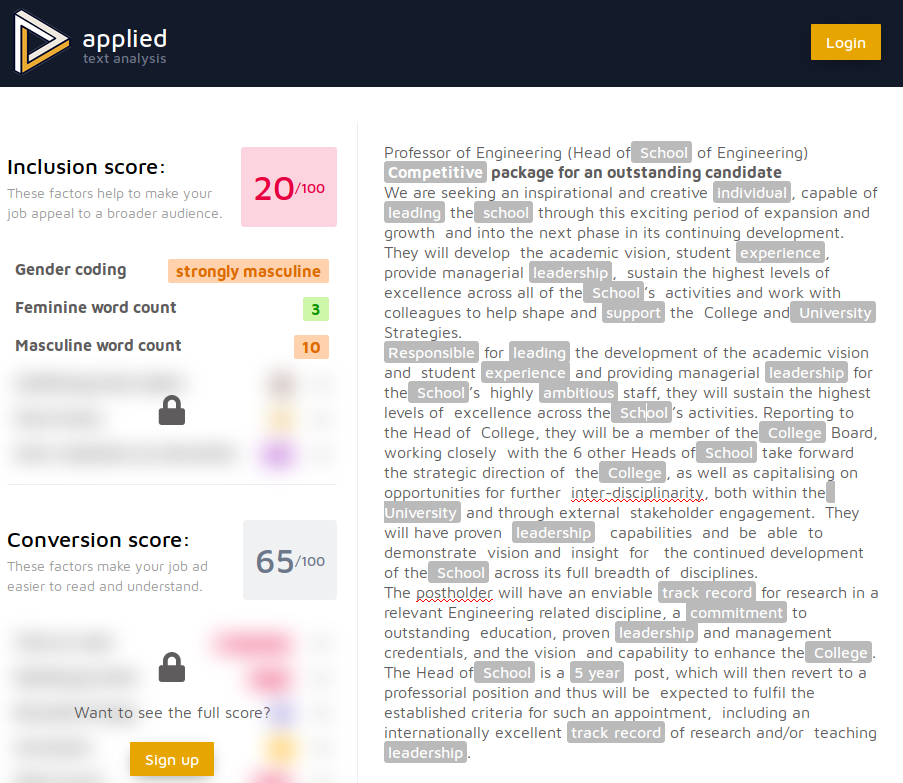}
  \caption{Example of gendered language and inclusion analysis of
  Applied. Generated with trial version available at \url{https://textanalysis.beapplied.com/} and an academic job ad of a British university.}
  \label{fig:applied-text}
\end{figure}

\begin{figure}[h]
  \centering
  \includegraphics[width=\linewidth]{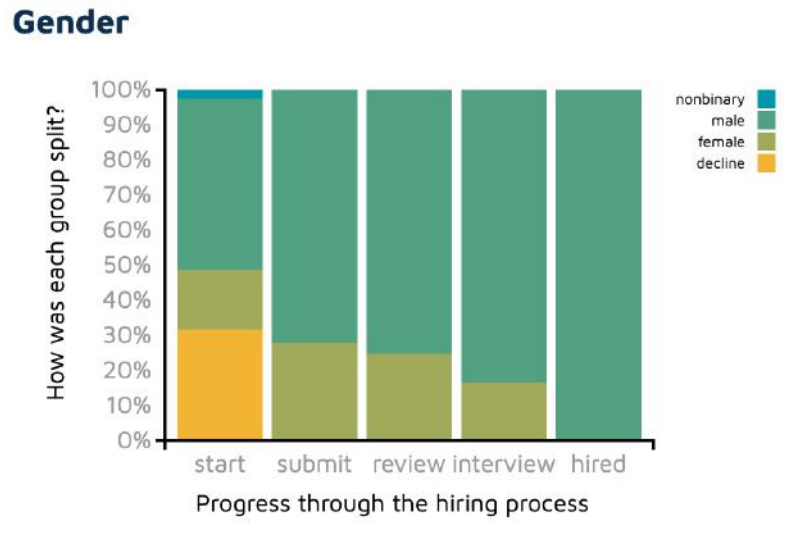}
  \caption{Visual auditing of bias implemented by Applied. Source
  \cite{applied_scaling_2019}.}
  \label{fig:applied-process}
\end{figure}

In contrast to Pymetrics and HireVue, it performs bias discovery and
mitigation by providing a de-biasing guide and demographic analytics
reports for the hiring pipeline \cite{applied_scaling_2019}. Note this tool does not perform
automatic candidate assessmet but it semi-automates the task of
discrimination monitoring. For example, regarding advertisement, the
platform can analyse gendered language use and inclusiveness of position
descriptions (see Figure \ref{fig:applied-text}). To `remove bias' in the rest of the steps,
the platform collects demographic information and then performs
candidate anonymization and removes direct and indirect group
information. Applied recommends to chunk assessment tests and compare
results across candidates rather than performing full reviews of
applications. In addition, it suggests to randomise the order of the
chunks and get more than one person to score each chunk. Rather than
performing formal statistical tests, the platform provides aggregated
analytics to evaluate the whole process to visually detect biases at
different stages (see Figure \ref{fig:applied-process}). Other comparison analytics are chunk
scoring for each group or the degree of scoring agreement between
multiple reviewers.

\section{Evaluating claims of bias mitigation}

Having outlined the different claims and approaches that our three AHSs
under study provide for mitigating bias in hiring we now draw on
FAT-related debates to briefly sketch some of the limitations of such
approaches.\footnote{This is not intended as an exhaustive list, but
  merely to highlight key concerns and debates.}

With different levels of formalization, the three AHSs understand
fairness in the broader categories identified in recent literature
\cite{corbett-davies_measure_2018}: (1) \emph{anti-classification}, the omission of group
variables and proxies in the decision making model, also known as
debiasing; (2) \emph{classification parity} in terms of equal passing
rates; and (3) \emph{calibration}, the requirement that outcomes are
independent of the group variables. Notice that all the categories
depend on clear definitions of groups. Examples of group definitions are
binary gender, ethnicity (using definitions informed specifically by US
demographics) and age interval. Social class or disabilities are not
found in either the examples nor in the available validation tests. (2)
and (3) depend on the comparison of an outcome that is ultimately the
selection of candidates, i.e. passing rates, but also candidate scoring.
HireVue and Pymetrics audit bias by checking the 4/5ths rule
(classification parity). Additionally, Pymetrics uses statistical tests
to compare group scores and passing rates which are a common way of
comparing groups represented by samples.

In line with on-going discussions on criteria for fairness in data
systems, notions of bias mitigation in AHSs present some inherent
limitations \cite{bogen_help_2018,raghavan_mitigating_2020}. One fundamental limitation is that reference
data is, by definition, extracted from current and past employees, and
in many cases from the `best performing' ones. Regardless of attention
to `fairness', the algorithmic specification of `best performing' or its
close association best `fit', can itself become a vehicle for bias
\cite{ajunwa_ifeoma_platforms_2019}. Within this setting, it is not clear that fair evaluation
methods can be built on past data that reflect historical injustices
\cite{bogen_help_2018,raghavan_mitigating_2020}. In other words, the issue with bias in AHSs may be in the
very logic of prediction itself. Related to this, if the data used to
build and validate the models only incorporates data from employees in
the company and not the rejected candidates, this limits the use of
fairness metrics that consider that discrimination can be better
reflected when accounting for disparate mistreatment \cite{zafar_fairness_2019}, for
example by comparing false negative cases, i.e., candidates that would
be suitable employees and were wrongly rejected. Validating fairness
with respect to passing rates (disparate impact), as in true positives,
has been discussed as problematic since it represents the degree of
belief we have in the model prediction with respect to an individual or
group rather than a measure of discrimination \cite{hellman_measuring_2019}. Hellman \cite{hellman_measuring_2019}
proposes to compare error ratios instead as a better measure to compare
different treatment of groups, which opens up a complex discussion about
the validity of metrics to compare groups.

Discrimination with respect to different attributes is partially covered
in the frameworks outlined by the three AHSs under study. Yet, as is a
common feature of bias mitigation, they appear limited to single-axis
understandings of identification that neglects substantial engagement
with intersectional forms of discrimination \cite{hoffmann_where_2019}. Pymetrics
validates fairness of each assessment test by performing multiple group
comparisons across each attribute, e.g. comparing its determined racial
groups scorings. Multiple axes of group identification are not directly
compared, e.g. hispanic women with white men. HireVue, meanwhile,
proposes in its patent to create a model that mitigates discrimination
by modifying a standard objective function to equally account for the
prediction errors of all the groups and also to penalize solutions that
violate the classification parity constraint. However, this proposal is
abstract and with no validation reports making it difficult to analyse.
Moreover, the idea of adding a set of penalty terms to the objective
function does not necessarily generate useful models. The problem arises
in the fact that adding many terms to these functions will decrease the
influence of each term, causing convergence problems in the learning
algorithm. Indeed, the study of intersectionality and rich subgroups
fairness in ML remains limited and in an early stage because of the
problem of multiple objectives for optimization and the complexity of
expressing the concept of intersectionality in a mathematical way
\cite{balashankar_pareto-efficient_2019,buolamwini_gender_2018,kearns_preventing_2018,zafar_fairness_2019}. In this, understandings of bias mitigation in AHSs
tend to overlook established limits of dominant antidiscrimination
discourses that have also featured in legal debates \cite{hoffmann_where_2019}.

These computational limitations of bias mitigation in AHSs point to
several directions that such efforts might take. However, none of these
discussions take account of the different national contexts in which
these systems are being deployed and how different legal frameworks
might apply, despite the prominence of legal definitions in the outline
of the system design. In order to illustrate the significance of this,
we now turn to discuss the relevant legal framework for the deployment
of such systems in a UK context where we know they are used, and how the
approaches to bias mitigation they propose relate.

\section{Legal framework}

Raghavan et al's \cite{raghavan_mitigating_2020} comprehensive study of automated hiring
systems (AHSs) makes it clear that the majority of such systems on the
market are developed in the US. Exceptions include Applied and Thrivemap
(both UK), Teamscope (Estonia) and ActiView (Israel). No rigorous
evidence seems to be available as to the market share of systems in
actual use across the globe, yet in terms of bias mitigation, where the
system is developed is crucial. While hiring goals and values may be
universal (itself questionable), legal regulation of data-driven
automated systems is decidedly not. As outlined above, considerations of
bias and its mitigation in US-built systems -- such as Pymetrics and
HireView - have clearly aimed at meeting the constraints of US equality
law \cite{barocas_solon_big_2016}, notably the 4/5ths rule set by the Uniform Guidelines
on Employee Selection Procedures by the Equal Employment Opportunity
Commission (EEOC) \cite{us_eeoc_adoption_1979}. Yet UK law (itself heavily permeated by EU
law) on equality and discrimination in employment matters is
considerably different; most obviously in the UK there is no 4/5ths
rule and hard statistical goals to (dis)prove bias do not seem formally
to exist, in either statute or case law. It is noticeable that the two
US-originated systems make use of the 4/5ths rule while the UK system
discussed (Applied) does not. Interestingly, also, another UK system
used widely, Thrive Map makes no claims as to bias mitigation at all and
is therefore not included in this paper.

Furthermore, EU DP law, now codified and reformed in the GDPR, provides
a suite of rights to ``data subjects'' (natural persons whose personal
data is processed) which are unknown to Federal US law (although
sectoral data rights do exist in the US in health, finance and some
other domains, and significant \emph{state} laws have been enacted eg
the Californian Consumer Privacy Act 2018, which replicates many
features of EU DP law). DP rights may prove in the UK/EU context to be
as or more important in uncovering or combating bias than sectoral
employment or equality rights; yet US systems may (unsurprisingly) not
be optimised to meet these rights.

Turning first to UK discrimination law, similarly to US law, it is based
initially on the idea of a closed list of ``protected characteristics'',
here laid out in the Equality Act 2010, s 4. UK law also however
(currently - Brexit may change this) has to respect supranational EU
law, which includes a number of Directives relevant to equality and
bias, notably the Equal Opportunities and Equal Treatment Directive
2006/54/EC, the Gender Equality Directive 2004/113/EC and the Framework
Directive for Equal Treatment in Employment and Occupation 2000/78/EC.
Human rights standards under the European Convention of Human Rights
(ECHR) are also relevant and refers to broader protection against
discrimination on \emph{any} ground (art 14) in relation to the rights
and freedoms guaranteed by the ECHR. The EU itself now has as a binding
source of law its own human rights instrument, the Charter of
Fundamental Rights of the EU \cite{noauthor_charter_2016}. This complex legislative
patchwork, divided across common law and civilian approaches to
law-making and interpretation \cite{lane_indirect_2018}, possibly contributes to fewer
``hard and fast'' standards for measuring unlawful bias in UK employment
law than in the US, which itself arguably makes the task of proving
debiasing harder for builders of AHSs for the UK market and may have
contributed to abandoning the effort altogether for some companies.

The UK's Equality Act 2010 attempts to replace a number of piecemeal
prior laws relating to different types of inequality with a coherent
statute covering inter alia sex, race and disability discrimination.
Rather as with US law, discrimination can be direct or indirect. Direct
discrimination in employment is nowadays regarded as rare \cite{sargeant_discrimination_2017}.
Indirect discrimination is defined in s 19(1) as ``a provision,
criterion or practice which is discriminatory in relation to a protected
characteristic''. Effectively it occurs when a policy that applies in
the same way to everybody has an effect that particularly disadvantages
people with a protected characteristic. This is similar to the US idea
of disparate impact. However the Code of Practice which accompanies the
Equality Act \cite{noauthor_equality_2011} does not lay down a statistical rule of thumb akin
to the 4/5ths rule to prove bias. Instead, reflecting s 19(2), the Code
provides only that a comparison must be made between workers with the
protected characteristic and those without it. The circumstances of the
two groups must be sufficiently similar for a comparison to be made and
there must be no material differences in circumstances (para 4.15). This
``pool for comparison'' consists of the group which the provision,
criterion or practice affects (or would affect) either positively or
negatively, while excluding workers who are not affected by it, either
positively or negatively. Importantly, the guidance does not always
require a formal comparative exercise using statistical evidence. Such
an approach was however adopted by the Court of Justice of the European
Union (CJEU) to prove indirect sex discrimination in \emph{R v Secretary
for State for Employment es parte Seymour-Smith} [2000] IRLR 263 and
is endorsed in some cases by the Equality Act Code for the UK as below
(para 4.21):

\begin{quote}
``• What proportion of the pool has the particular protected
characteristic?

• Within the pool, does the provision, criterion or practice affect
workers

without the protected characteristic?

• How many of these workers are (or would be) disadvantaged by it?

How is this expressed as a proportion (`x')?

• Within the pool, how does the provision, criterion or practice affect

people who share the protected characteristic?

• How many of these workers are (or would be) put at a disadvantage by
it?

How is this expressed as a proportion (`y')?

Using this approach, the Employment Tribunal will then compare (x) with
(y).''
\end{quote}

However there is no particular prescribed ratio of outcomes that proves
bias. ``Whether a difference is significant will depend on the context,
such as the size of the pool and the numbers behind the proportions. It
is not necessary to show that the majority of those within the pool who
share the protected characteristic are placed at a ``disadvantage''
(para 4.22). Furthermore according to s 19(2), bias can be justified if
it is shown to be a ``proportionate'' means of achieving a
``legitimate'' aim. Legitimacy is not even defined in the 2010 Act
though guidance can be drawn from the CJEU and the Code of Practice,
which states that the aim of the discriminatory provision, criterion or
practice ``should be legal, should not be discriminatory in itself, and
must represent a real, objective consideration''. It adds: ``Although
reasonable business needs and economic efficiency may be legitimate
aims, an employer solely aiming to reduce costs cannot expect to satisfy
the test'' (para 4.29). Given this kind of language, it is hard to
imagine how either proportionality or legitimacy could be coded into a
hiring tool. In legal discourse, proportionality is what is known as an
``open textured'' concept: it is impossible to predict what factors will
come relevantly into play in advance, lacking a sufficiently large
dataset of case law, nor how factors would be ranked. Even if sufficient
data was available to mine using ML techniques, we would argue that it
would not contain the individualised policy factors which drive courts
or tribunals to make decisions around proportionality and legitimacy. by
contrast a hard edged heuristic like the 4/5ths rule is simplistic to
implement.

Turning to DP law, all data subjects have a number of rights in relation
to the processing of personal data (itself defined in GDPR, art 4(1)),
including rights to transparency and access to data held about them
(``subject access rights'' or SARs)(GDPR, arts 12-15) and to object to
decisions which have legal or similarly significant effects being made
about them using their personal data and by solely automated means
(GDPR, art 22). The latter provision caused great academic stir in 2016
when it was claimed somewhat controversially it could be interpreted to
provide data subjects, not just with a right to a ``human in the loop''
as had been known to be the case (albeit with little publicity) since
1995, but also with a ``right to an explanation'' of how the decision
was made \cite{edwards_slave_2017,goodman_european_2017,wachter_why_2017}. Given the importance of access to employment,
automated hiring systems almost certainly make a decision which has
legal or significant effect. Indeed, the Information Commissioner's
Office (ICO), which regulates DP in the UK, gives ``e-recruiting
practices without human intervention'' as a canonical example for the
application of art 22\footnote{See ICO guidance on the GDPR and DPA 2018
  at
  \href{https://ico.org.uk/for-organisations/guide-to-data-protection/guide-to-the-general-data-protection-regulation-gdpr/individual-rights/rights-related-to-automated-decision-making-including-profiling/}{{https://ico.org.uk/for-organisations/guide-to-data-protection/guide-to-the-general-data-protection-regulation-gdpr/individual-rights/rights-related-to-automated-decision-making-including-profiling/}},}.

Thus it could be argued that any use of a fully automated AHS,
regardless of whether bias can be proved or is, indeed, said to have
been mitigated, can be refused by the prospective employee on the ground
that it is a solely automated decision under art 22, requiring explicit
consent (art 22(2)(c)) and instead, the candidate could ask for a human
to make that decision instead, or reconsider it. This rule is augmented
further by the fact that it may be impossible to give valid explicit
consent in the context of employment anyhow. Consent under the GDPR art
4(11) must be ``freely given'' and the Art 29 Working Party (A29 WP -
now replaced by the European Data Protection Board or EDPB) who provide
persuasive guidance on the GDPR, have indicated that truly free, and
therefore valid, consent can probably never be given in the context of
employment relations \cite{noauthor_article_2016}. Of course it could be argued that a
hiring (though not a firing or promotion/demotion) system
\emph{precedes} employment, and consent can therefore be valid; it seems
unlikely the CJEU would take kindly to this, especially in times of
austerity and precarity. Thus by chain of deduction it seems plausible
that it is in fact illegal to use a solely automated AHS in the EU.

Issues arise with this analysis however. First, anecdotally, very few
hiring (or even firing, promotion, demotion, allocation of hours, etc.)
decisions seem to be taken without any human intervention at all. To
take a recent US example, Amazon came under fire in April 2019 for
apparently automatedly sacking up to 10\% a year of their employees
whose productivity fell below certain measured efficiency levels in
environments regarded as highly datafied \cite{lecher_how_2019}. However later
evidence emerged that no ``automatic'' sackings in fact occurred and a
human supervisor was always there to reverse the sacking. Thus art 22
would arguably not have applied. Uber's global driver terms and
conditions state that automated firings can take place but then add that
in the EU a right to object to a human is available \cite{uber_technologies_inc._uber_2019}. This
suggests however that mere rubber stamping by a manager with no real
intervention into decision making might be sufficient to render art 22
nugatory. What constitutes ``enough'' interaction by a human such that
art 22 is not triggered remains an unclear issue in the GDPR, and may
vary from member state to member state (see \cite{veale_clarity_2018,noauthor_article_2016}). Secondly, art
22 does not require consent from the data subject if the decision is
``necessary for entering into or performance of a contract between an
organisation and the individual'' (art 22 (1)(a))\footnote{A second
  exemption relates to where the decision is authorised by member state
  or EU law. This refers to governmental ``public tasks'' and it seems
  unlikely it could ever apply to an AHS.}. Could submission to an AHS
ever be ``necessary'' for entering the contract of employment? It seems
prima facie unlikely but an employer might argue eg triage when many
1000s of applications are received does not just benefit from but
actually \emph{requires} solely automated systems.

If the right to object under art 22 is excluded by the lack of a
``solely automated'' decision, then any ``right to an explanation'' read
from art 22 may also fall. However it is possible, though also
controversial, that such a right may then be derived from art 15(h)
which provides that users have a right to information about ``the
existence of automated decision-making, including profiling'' (so the
use of an AHS in hiring has to be notified to candidates) and ``at least
in those cases, meaningful information about the logic involved''
\cite{veale_clarity_2018}. It can be argued that the use of the phrase ``at least'' means
that semi-automated decision making may not exclude the right to
``meaningful information'' \cite{veale_clarity_2018}. Again such a right need not be
chained to proof of bias, or failure to mitigate bias, and yet could
prove highly effective in exposing discriminatory or even simply
arbitrary or erroneous practices, at an individual and possibly even at
a group level, given the possibility for collective redress actions
within the GDPR (see arts 80 and 82). Thus at this point an easier route
to disincentivising bias in AHSs might, in the EU context, be seen as
coming via DP rather than equality rights - especially given the
probable difficulty of building in bias mitigation into AHSs
definitively capable of meeting EU legal standards. It is also worth
noting here that any machine learning system is likely to be regarded as
``high risk'' processing requiring a prior Data Protection Impact
Assessment (DPIA) (see art 35(3)(a), which should show inter alia that
potential for unfairness and bias had been considered and steps taken to
avert. This might arguably be seen as implying a \emph{requirement} for
debiasing in AHS tools deployed in EU (see also \cite{ajunwa_paradox_2020} for a similar
point in a US context).

However, even if a right to algorithmic transparency does exist in the
solely-automated hiring context, what does it practically \emph{mean}?
This has again been the subject of much academic debate. Selbst and
Powles argue, for example, that for a right to ``meaningful information
about the logic involved'' to be (sic) meaningful, it must be more than
a simple regurgitation of source code \cite{selbst_meaningful_2017}. The A29 Working Party
recommend that the data subject should be provided with ``general
information (notably, on factors taken into account for the
decision-making process, and on their respective `weight' on an
aggregate level) which is also useful for him or her to challenge the
decision'' \cite{cappelli_data_2019}. To date there has been no relevant case law in the
UK and the provision is not expanded on in the Data Protection Act (DPA)
2018 which implements the GDPR in the UK (in comparison to some other
member states such as France).

\section{Discussion}

The computational and legal challenges of AHSs that we have outlined
here raise significant concerns for tackling discrimination and
providing transparency to enable challenges to AHS hiring decisions in
the context of the UK. This is particularly pertinent as bias mitigation
in hiring is one of the key selling points for several of these tools.
They are part of a growing `diversity, equity, and inclusion' (D.E.I)
industry that has boomed in the last couple of years \cite{zelevansky_big_2019}. Whilst a few of
the companies developing AHSs provide some documentation to evidence
such claims, access to relevant information remains a key problem for
conducting any thorough analysis. Claims and validation are often vague
and abstract, if they are provided at all. Moreover, it is not clear how
relevant stakeholders, not least job seekers, are able to access and
understand information about how decisions about their eligibility might
have been reached through AHSs. This makes it difficult to assess if and
how discriminatory practices might have been part of the hiring process,
and leaves little room for anyone to challenge the decision made. Given
the transparency rights attributed to data subjects by the GDPR, this
haziness as to transparency is unacceptable in the EU and UK. On the
other hand, what approach to transparency is required by EU law, remains
itself vague. It would be good to see AHSs built in, or sold into the EU
market meeting this challenge of ``meaningful information'' explicitly.
Might it take the form of the counterfactual explanations which have
become fashionable but actually offer little by way of practical
remedies? \cite{wachter_counterfactual_2018} What if vendors say that greater transparency is
simply not possible? Again we may be back at a conclusion that such
systems simply cannot be lawful in the EU.

The AHSs we have looked at in this paper are relatively unique in
providing some information about their workings, even if the exact data
sources and model remain obscure and will vary according to different
clients. In particular, in seeking to explicitly tackle issues of
discrimination in hiring practices, these systems provide some insights
into how such issues are understood and approached by AHSs. This is
significant, and welcomed, as it provides an opportunity to engage with
what could be considered as emerging standards in managerial techniques
that are being exported to a global marketplace.

Whilst a desire to address the prevalence of bias and discrimination in
hiring is a significant pursuit (not least in the context of the UK
where levels of discrimination against ethnic minorities in accessing
jobs have remained relatively unchanged since the 1960s \cite{siddique_minority_2019}), the
approaches to bias mitigation provided by the three AHSs we have looked
at come with important \textbf{limitations.} Here, we are not attempting
to provide a comprehensive list of the issues that might come with the
use of AHSs in general, but want to summarise a few of the key points
that emerge from the evidence base we have provided in this paper.
First, attempts at mitigation within AHSs run into on-going concerns
with computational \emph{fairness}. These relate not only to the
inherent problems with data-driven predictions and with relying on
quantification for determining criteria of `good' or appropriate `fit',
but also to the necessary reductionist nature of group identification in
computational systems and the neglect of intersectionality that have
also been the subject of significant criticism in legal understandings
of discrimination.

Secondly, attempts at bias mitigation in AHSs within a UK context also
show problems with \emph{accountability}. When such technologies are
used for decision-making, to whom are the companies making AHSs
responsible? The employer or the candidate? The employer might argue
that some types of transparency are sometimes undesirable, as indeed
might the software company defending its intellectual property. Thus,
the question is whose job it is to fulfil the obligation of transparency
alongside the obligation of bias mitigation; the system builder, the
employer who utilises it, or another actor altogether.

Connectedly , the transfer of AHSs developed within a US socio-legal
context to a UK (and arguably EU) context introduces a number of
fundamental legal problems of fit, not just with regards to
discrimination and equality law, but as we have argued, perhaps more
significantly in relation to DP law. GDPR transparency rights in arts 15
and 22 may provide avenues to overturn aspects of the candidate-employer
information asymmetry and might even outright prohibit the use of AHSs
for wholly automated decision-making in hiring. Yet these rights may be
ignored or ill-implemented in systems not built within the EU.

A number of other points might be addressed in \textbf{future work}. US
literature on bias, especially racial bias, in the algorithmic workplace
focuses on hiring because the US in general has ``at will'' firing with
few legal constraints \cite{ajunwa_limitless_2016}. In Europe and in the UK specifically,
things are very different and a quick survey of Employment Appeal
tribunal cases shows most revolve around firing or issues of in-work
conditions. There is a real need in extending FAT work on algorithmic
bias in the workplace to Europe to consider these other loci for
datafication, bias and opacity. Furthermore, more work is needed looking
specifically at systems developed in the UK and the EU that also
connects these to the actual practices and experiences of employers and
candidates to get a sense of how AHSs shape those interactions. Such
work requires more qualitative research that we seek to pursue in future
project work.

In conclusion, the lack of information about how AHSs work, the approach
they take to tackling discriminatory hiring practices, and crucially,
where and how they are used around the world is therefore a significant
problem. Given it is not even clear that AHSs provide significant
benefits to employers \cite{cappelli_data_2019}, it could even be asked if their use
should actually be restricted or discouraged by regulators from the EU
data protection and equality sectors. As this trend is set to become
more pervasive, there is an urgent need to assess if and how AHSs should
be used so as to uphold fundamental rights and protect the interests of
candidates and employees.

\begin{acks}
  The research of Lina Dencik and Javier Sánchez-Monedero has been funded
by the ERC Starting Grant DATAJUSTICE (grant no. 759903).
\end{acks}

\bibliographystyle{ACM-Reference-Format}
\bibliography{references.bib}

\end{document}